\documentclass[11pt]{article}
\usepackage{epstopdf}
\usepackage{color}
\usepackage{subfigure}
\usepackage{amsmath}
\usepackage{amssymb}
\usepackage{graphicx,color}
\usepackage{cite}
\usepackage{enumerate}
\usepackage{amsthm}
\usepackage{amsfonts,mathrsfs}
\usepackage{geometry}
\usepackage{ifthen}

\begin{document}

\title{\bf Average skew information-based coherence and its typicality for random quantum states}

\vskip0.1in
\author{\small Zhaoqi Wu$^{1,4}$, Lin Zhang$^{2,4}$\thanks{Corresponding author. E-mail:godyalin@163.com},
Shao-Ming Fei$^{3,4}$,
Xianqing Li-Jost$^{4}$\\
{\small\it  1. Department of Mathematics, Nanchang University, Nanchang 330031, P R China} \\
{\small\it  2. Institute of Mathematics, Hangzhou Dianzi University, Hangzhou 310018, P R China}\\
{\small\it  3. School of Mathematical Sciences, Capital Normal University, Beijing 100048, P R China}\\
{\small\it  4. Max Planck Institute for Mathematics in the Sciences,
04103 Leipzig, Germany}}

\date{}
\maketitle

\noindent {\bf Abstract} {\small } We study the average skew
information-based coherence for both random pure and mixed states.
The explicit formulae of the average skew information-based
coherence are derived and shown to be the functions of the dimension
$N$ of the state space. We demonstrate that as $N$ approaches to
infinity, the average coherence is $1$ for random pure states, and a
positive constant less than 1/2 for random mixed states. We also
explore the typicality of average skew information-based coherence
of random quantum states. Furthermore, we identify a coherent
subspace such that the amount of the skew information-based
coherence for each pure state in this subspace can be bounded from
below almost always by a fixed number that is arbitrarily close to
the typical value of coherence.

\vskip 0.1in


\vskip0.2in

\noindent {\bf 1. Introduction}

\vskip0.1in

Quantum coherence is a fundamental issue in quantum mechanics, and
an important physical resource in quantum information theory
\cite{STRELTSOV}. An axiomatic definition of a valid quantum
coherence measure has been proposed in \cite{TB}, which intrigued
great interest in quantifying and studying the properties of quantum
coherence. Many distance measures and information related
quantities, such as relative entropy \cite{TB}, $l_1$ norm
\cite{TB}, robustness of coherence \cite{NAPOLI}, max-relative
entropy \cite{KB1}, geometric coherence \cite{AS1,CX1}, fidelity
\cite{LHS}, trace distance \cite{RANA}, modified trace distance
\cite{XDY,CHEN}, skew information \cite{CSY,LUO1,LUO2,LUO3},
coherence weight \cite{KB2}, affinity \cite{CX2,CX3}, generalized
$\alpha$-$z$-relative R\'enyi entropy \cite{XNZ} and logarithmic
coherence number \cite{ZJX}, have been exploited to quantify quantum
coherence. Quantum coherence from other resource-theoretical
perspectives, such as coherence distillation and coherence dilution
\cite{AW,EC2,BR,YUAN3,KF,CLL,LL,MJZ,QZ}, no-broadcasting of quantum
coherence \cite{MLMPM,IMRWS}, interconversion between quantum
coherence and quantum entanglement \cite{AS1,EC1,HJZ,YX} or quantum
correlations \cite{JM,LUO5,MLH,KIM,KDW,GUO} and cohering power of
quantum operations \cite{KB3}. Coherence manipulation under
incoherent operations \cite{SDU} have also been studied.

Wigner-Yanase (WY) skew information \cite{WY} is a very important
information quantity, which has been widely used and explored in
studying quantum information problems in recent years. WY skew
information has been exploited to define different coherence
measures, such as $K$-coherence \cite{GIROLAMI}, modified
$K$-coherence \cite{LUO2} and skew information-based coherence
\cite{CSY}. In particular, skew information-based coherence was
proven to be a well-defined measure, with tight connections with
quantum correlations and the corresponding experimental
implementations.

In quantifying the coherence of a quantum state \cite{TB}, a
coherence measure is defined with respect to a certain basis. To
eliminate the impact of the basis, two questions need to be
addressed: first of all, for a given coherence measure, if the
coherence of a state with respect to one basis is very large, how
large the coherence of it could be with respect to another basis?
Does any tradeoff relation exists? This question has been examined
for a set of mutually unbiased bases (MUBs) for $l_1$ norm of
coherence and relative entropy of coherence in \cite{SC} and for
skew information-based coherence in \cite{LUO4}. Secondly, is it
possible to characterize the coherence of a quantum state without
referring to any particular basis? This question is answered by
considering average measure of coherence over all bases \cite{SC}.
Since all reference bases can be generated from unitary operations
on a given basis, what we need is to calculate the integration over
the unitary orbit of a fixed basis, or equivalently, the integration
over the unitary group equipped with the normalized Haar measure
\cite{LUO4}. This averaging shows the degree to which the state is
coherent if a basis is chosen at random, which has been studied for
$l_1$ norm of coherence and relative entropy of coherence in
\cite{SC} and for skew information-based coherence in \cite{LUO4}.
It is found that for skew information-based coherence, the average
coherence over all orthonormal bases is equal to the average
coherence over any complete MUBs. These study reveals intrinsic
essence of coherence encoded in a state.

On the other hand, the random matrix theory provides new
perspectives to study quantum physics and quantum information theory
\cite{COLLINS}. From the view of probability and statistics, average
value represents the first moment, which is an important numerical
characteristics, and can further characterize some problems such as
the law of large numbers and other convergence properties. Random
pure quantum states possess many important properties including the
concentration of measure phenomenon or typicality \cite{ML}, which
allow one to get more information on the structures of the quantum
system \cite{COLLINS,ML,HAYDEN1,HAYDEN2}. The entanglement features
of pure bipartite quantum states sampled from the uniform Haar
measure have been studied in recent years
\cite{HAYDEN1,HAYDEN2,DNP,SKF,JSR,SEN,LCM,AD,AH,OCOD,LZ1,RFW}, among
which the average entropy of a subsystem has been calculated and
investigated \cite{DNP,SKF,JSR}. It has been shown that a typical
pure state of an $N\times N$ system is almost maximally entangled
\cite{HAYDEN2}. New analytical formulae describing the levels of
entanglement expected in random pure states have also been presented
\cite{AJS}.

The results in \cite{SC} and \cite{LUO4} concern the average
coherence of given quantum states. It is thus natural to consider
average coherence of random quantum states with respect to the Haar
measure on the unitary group. Based on the average value of
coherence, the concentration of measure phenomenon can be further
studied, which can reveal statistical behavior and characteristics
of quantum coherence. In recent years, average coherence based on
relative entropy of coherence and its typicality for random pure
states \cite{LZ2} and random mixed states \cite{LZ3} have been
derived, and average subentropy, coherence and entanglement of
random mixed quantum states have been discussed \cite{LZ4}.
Moreover, the average of uncertainty product for bounded observables
has been also calculated \cite{LZ5}.

Since skew information-based coherence is of great significance, the
following questions naturally arise: can we calculate the average
skew information-based coherence for random pure/mixed states? what
is the concentration measure of phenomenon (typicality) of this
average coherence for random pure/mixed states? In this paper, we
will answer these questions.

The paper is organized as follows. We begin with a retrospect of the
framework for quantification of coherence, skew information and the
coherence measure based on it in Sec. 2. In Sec. 3, we recall random
pure quantum states, L\'evy's Lemma, random mixed quantum states and
related preliminaries. In Sec. 4, we calculate the average skew
information-based coherence for random pure states sampled from the
uniform Haar distribution, investigate the typicality of the
obtained average coherence, and figure out the dimension of the
subspace of the total Hilbert space such that all the pure states in
this subspace have a fixed nonzero amount of coherence. For random
mixed states, we also calculate the average skew information-based
coherence and study its typicality in Sec. 5, which turned out to
have different features compared with random pure states. Finally,
we conclude in Sec. 6 with a summary and discussions on the
significance and implementations of the obtained results.

\vskip0.1in

\noindent {\bf 2. Skew information-based coherence}

\vskip0.1in

Let $\mathcal{H}=\mathbb{C}^N$ be a Hilbert space of dimension $N$,
and $\mathrm{B}\mathcal{(H)}$, $\mathrm{S}\mathcal{(H)}$ and
$\mathrm{D}\mathcal{(H)}$ be the set of all bounded linear
operators, Hermitian operators and density operators on
$\mathcal{H}$, respectively. Mathematically, a state and a channel are
 described by a density operator (positive operator of
trace $1$) and a completely positive trace preserving (CPTP) map,
respectively \cite{NC}.

Fix an orthonormal basis $\{|k\rangle\}^N_{k=1}$ of $\mathcal{H}$.
The set of incoherent states, which are diagonal in this basis, can
be written as
$$\mathcal{I}=\{\delta\in \mathrm{D}\mathcal{(H)}|\delta=\sum^N_{k=1}p_k|k\rangle\langle k|,~p_k\geq 0,~\sum^N_{k}p_k=1\}.$$
Let $\Lambda$ be a CPTP map $\Lambda(\rho)=\sum_{n}K_n\rho
K_n^\dag,$ where $K_n$ are Kraus operators satisfying
$\sum_{n}K_n^\dag K_n=I_{N}$ with $I_N$ the identity operator on
$\mathcal{H}$. $K_n$ are called incoherent Kraus operators if
$K_n^\dag \mathcal{I}K_n\in \mathcal{I}$ for all $n$, and the
corresponding $\Lambda$ is called an incoherent operation.

A well-defined coherence measure $C(\rho)$ of a quantum state should
satisfy the following conditions \cite{TB}:

$(C1)$ (Faithfulness) $C(\rho)\geq 0$ and $C(\rho)=0$ iff $\rho$ is
incoherent.

$(C2)$ (Convexity) $C(\cdot)$ is convex in $\rho$.

$(C3)$ (Monotonicity) $C(\Lambda(\rho))\leq C(\rho)$ for any
incoherent operation $\Lambda$.

$(C4)$ (Strong monotonicity) $C(\cdot)$ does not increase on average
under selective incoherent operations, i.e., $C(\rho)\geq
\sum_{n}p_nC(\varrho_n),$ where $p_n=\mathrm{Tr}(K_n\rho K_n^\dag)$
are probabilities and $\varrho_n=\frac{K_n\rho K_n^\dag}{p_n}$ are
the post-measurement states, $K_n$ are incoherent Kraus operators.

For a state $\rho\in \mathrm{D}\mathcal{(H)}$ and an observable
$K\in \mathrm{S}\mathcal{(H)}$, the {\it Wigner-Yanase} (WY) skew
information is defined by \cite{WY}
\begin{equation}\label{eq1}
I(\rho,K)=-\frac{1}{2}\mathrm{Tr}([\rho^{\frac{1}{2}},K]^2),
\end{equation}
where $[X,Y]:=XY-YX$ is the commutator of $X$ and $Y$.

Girolami has utilized the Wigner-Yanase skew information $I(\rho,K)$
to give a coherence measure in a direct manner, where $K$ is
diagonal in the basis $\{|k\rangle\}_{k=1}^N$, and called it
$K$-coherence \cite{GIROLAMI}. This quantity is in
fact a quantifier for coherence of $\rho$ with respect to $K$ rather
than the orthonormal basis $\{|k\rangle\}_{k=1}^N$.

It is argued that the $K$-coherence satisfies $(C1)$ and $(C2)$, but
fails to meet the requirement $(C3)$ \cite{DUBAI,MARVIAN}. By
considering coherence with respect to the L\"uders measurements
induced from the observable $K$, it is shown that the $K$-coherence
can be readily adapted to a bona fide measure of coherence
satisfying $(C1)$-$(C3)$ \cite{LUO2} (which is the coherence in the
context of partially decoherent operations, and has been called
partial coherence in \cite{LUO2}). Another way to resolve the above
problem is to introduce the skew information-based coherence measure
defined by \cite{CSY}
\begin{equation}\label{eq2}
C_I(\rho)=\sum_{k=1}^N I(\rho,|k\rangle\langle k|),
\end{equation}
where $I(\rho,|k\rangle\langle
k|)=-\frac{1}{2}\mathrm{Tr}\{[\rho,|k\rangle \langle k|]\}^2$ is the
skew information of the state $\rho$ with respect to the projection
$|k\rangle \langle k|$. Direct calculations show that (\ref{eq2})
can be further written as \cite{CSY}
\begin{equation}\label{eq3}
C_I(\rho)=1-\sum_{k=1}^N \langle k|\sqrt{\rho}|k\rangle ^2.
\end{equation}
It is easy to see that $\max_{\rho}C_I(\rho)= 1-\frac{1}{N}$, and
the maximum is attained for the maximally coherent state
$|\psi\rangle=\frac{1}{\sqrt{N}}\sum_{j=1}^N
e^{i\theta_j}|j\rangle$.

If $\rho=|\psi\rangle \langle\psi|$ is a pure state, one has
\begin{equation}\label{eq4}
C_{I}(\psi)=1-\sum_{k=1}^N |\langle k|\psi\rangle|^4.
\end{equation}

In \cite{CSY}, it has been proved that the coherence measure defined
in (\ref{eq2}) satisfies all the criteria $(C1)$-$(C4)$, while the
$K$-coherence does not satisfy $(C4)$ (strong monotonicity). The
advantage of this coherence measure is that it has an analytic
expression. Also, an operational meaning in connection with quantum
metrology has been revealed. The distribution of this coherence
measure among the multipartite systems has been investigated and a
corresponding polygamy relation has been proposed. It is also found
that this coherence measure provides the natural upper bounds of
quantum correlations prepared by incoherent operations. Moreover, it
is shown that this coherence measure can be experimentally measured
\cite{CSY}. Since the skew information-based coherence measure
(\ref{eq2}) is well-defined and can be analytically expressed, it is of
great significance both theoretically and practically, and worth
evaluating the average coherence based on this measure for both
random pure quantum states and random mixed quantum states.

\vskip0.1in

\noindent {\bf 3. Random pure quantum states, L\'evy's Lemma, random
mixed quantum states}

\vskip0.1in

{\it Random pure quantum states.} Let $\mathcal{H}=\mathbb{C}^N$ be
a Hilbert space of dimension $N$, $\mathrm{U(N)}$ be the group of
all $N\times N$ unitary matrices, $\mathrm{M_{N}}(\mathbb{C})$ be
the set of all $N\times N$ complex matrices, and
$\mathrm{D}(\mathbb{C}^N)$ be the set of all density matrices on
$\mathbb{C}^N$. The set of pure states on $\mathbb{C}^N$ is the
complex projective space $\mathbb{C}\mathrm{P}^{N-1}$. For the space
of pure states $|\psi\rangle$ there exists a unique measure
$\mathrm{d}(\psi)$ induced from the uniform Haar measure
$\mathrm{d}\mu(U)$ on the unitary group $\mathrm{U(N)}$, which
implies that any random pure state $|\psi\rangle$ can be obtained
via a unitary operation on a fixed pure state $|\psi_0\rangle$:
$|\psi\rangle=U|\psi_0\rangle$. The average value of a function
$g(\psi)$ of pure states $|\psi\rangle$ is defined as
$$\mathbb{E}_{\psi}[g(\psi)]=\int \mathrm{d}(\psi)g(\psi)=\int_{\mathrm{U(N)}} \mathrm{d}\mu(U)g(U\psi_0).$$

{\it Lipschitz continuous function and Lipschitz constant.} Let
$(X,d_1)$ and $(Y,d_2)$ be two metric spaces and $T:X\rightarrow Y$
be a mapping. $T$ is called a Lipschitz continuous mapping on $X$
with the Lipschitz constant $\eta$, if there exists $\eta>0$ such
that
$$d_2(T(x),T(y))\leq \eta d_1(x,y)$$
holds for all $x,y\in X$ \cite{MO}. Note that any real number larger
than $\eta$ is also a Lipschitz constant for the mapping $T$
\cite{MO}.

In this work, we will use the concept of a Hilbert-Schmidt norm of a
matrix $A$, which is defined as $\|A\|_2:=\sqrt{\mathrm{Tr}A^\dag
A}$ \cite{MMW}. Also, in deriving the Lipschitz constant for
discussing the typicality for random pure/mixed states, we need the
notion of the gradient of a function. The best linear approximation
to a differentiable function $f:\mathbb{R}^n\rightarrow \mathbb{R}$
at a point $x$ in $\mathbb{R}^n$ is linear mapping from
$\mathbb{R}^n$ to $\mathbb{R}$ which is often denoted by
$\mathrm{d}f_x$ or $Df(x)$ and called the differential or (total)
derivative of $f$ at $x$. The \emph{gradient} is then related to the
differential by the formula $(\nabla f)_x\cdot v=\mathrm{d}f_x(v)$
for any $v\in \mathbb{R}^n$, that is, the one-form (i.e., a linear
functional) acting on vectors induced a vector
representation\footnote{This fact is just like Riesz representation
in Hilbert space.} $(\nabla f)_x$ with respect to the scalar
product. The function $\mathrm{d}f$, which maps $x$ to
$\mathrm{d}f_x$, is called the differential or (exterior) derivative
of $f$ and is an example of differential one-form. If $\mathbb{R}^n$
is viewed as the space of (dimension $n$) column vectors (of real
numbers), then one can regard $\mathrm{d}f$ as the row vector with
components $\left(\frac{\partial f}{\partial
x_1},\cdots,\frac{\partial f}{\partial x_n}\right)$, so that
$\mathrm{d}f_x(v)$ is given by matrix multiplication. The gradient
is then the corresponding column vector, i.e., $(\nabla
f)_i=\mathrm{d}f_i^{T}$ \cite{GAKTMK}.

{\it L\'evy's Lemma (see \cite{ML} and \cite{HAYDEN2})}. Let
$T:\mathbb{S}^k\rightarrow \mathbb{R}$ be a Lipschitz continuous
function from the $k$-sphere to the real line with a Lipschitz
constant $\eta$ (with respect to the Hilbert-Schmidt norm). Let
$z\in \mathbb{S}^k$ be a  chosen uniformly at random. Then for any
$\epsilon>0$, we have
\begin{equation}\label{eq5}
\mathrm{Pr}\{|T(z)-\mathbb{E}[T]|>\epsilon\}\leq
2\mathrm{exp}\left(-\frac{(k+1)\epsilon^2}{9\pi^3\eta^2\mathrm{ln2}}\right),
\end{equation}
where $\mathbb{E}[T]$ is the expected value of $T$.

Note that the average over the Haar distributed $N$-dimensional pure
states is equivalent to the average over the $k$ sphere with
$k=2N-1$.

{\it Existence of small nets.} To prove the existence of
concentrated subspaces with a fixed amount of coherence, we need the
notion of small nets \cite{HAYDEN1}. Given a Hilbert space
$\mathcal{H}$ of dimension $N$ and $0<\epsilon_0<1$, there exists a
set $\mathcal{N}$ of pure states in $\mathcal{H}$ with
$|\mathcal{N}|\leq (5/\epsilon_0)^{2N}$ such that for every pure
state $|\psi\rangle \in \mathcal{H}$, there exists
$|\tilde{\psi}\rangle \in \mathcal{N}$ such that
$\||\psi\rangle-|\tilde{\psi}\rangle\|_2\leq \frac{\epsilon_0}{2}$,
where $\|\cdot\|_2$ is the Hilbert-Schmidt norm of a matrix. This
set $\mathcal{N}$ is called an $\epsilon_0$ net.

{\it Random mixed quantum states.} Quantum ensembles are defined by
choosing probability measures on $\mathrm{D}(\mathbb{C}^N)$. It is
worth noting that such measure may not be unique, and different
measures may have different physical motivations, advantages and
drawbacks, while the Fubini-Study (FS) measure is the only natural
measure in defining random pure states \cite{IBKZ}.

We have to pay a high price for considering a Riemannian geometry on
$\mathrm{D}(\mathbb{C}^N)$, since it is usually difficult to tackle
with the emerged monotone metrics when $N>2$. Luckily, the measures
that induced from some chosen monotone metrics are not that
difficult to deal with. The technique is the same as that one uses
in flat space, when the Euclidean measure is decomposed into a
product. The set of quantum mixed states that can be written in the
form $\rho=U\Lambda U^\dagger$, for a fixed diagonal matrix
$\Lambda$ with strictly positive eigenvalues, is a flag manifold
${\bf F}^{(N)}=\mathrm{U(N)/[U(1)]}^N$. It is naturally assumed that
a probability distribution in $\mathrm{D}(\mathbb{C}^N)$ possess the
invariance with respect to unitary rotations, $P(\rho)=P(W\rho
W^\dagger)$. This assumption can be guaranteed if (a) the chosen
eigenvalues and eigenvectors are independent, and (b) the
eigenvectors are drawn according to the Haar measure,
$\mathrm{d\mu_{\mathrm{Haar}}}(W)=\mathrm{d\mu_{\mathrm{Haar}}}(UW)$
\cite{IBKZ}.

Combining the two measures, a product measure on the Cartesian
product of the flag manifold and the simplex ${\bf F}^{(N)}\times
\Delta_{N-1}$ can be defined: $\mathrm{d\mu(\rho)=
d\mu_{Haar}}(U)\times \mathrm{d}\nu(\Lambda)$, which induces the
corresponding probability distribution,
$P(\rho)=P_{\mathrm{Haar}}({\bf F}^{(N)})\times P(\Lambda)$, where
the first factor denotes the natural, unitarily invariant
distribution on the flag manifold ${\bf
F}^{(N)}=\mathrm{U(N)/[U(1)]}^N$ induced by the Haar measure on
$\mathrm{U(N)}$. Note that the Haar measure on $\mathrm{U(N)}$ is
unique while there is no unique choice for $\nu$ \cite{LZ4,IBKZ,
KZHJS}.

The measures used frequently over $\mathrm{D}(\mathbb{C}^N)$ can be
obtained by taking partial trace over a $M$-dimensional environment
of an ensemble of pure states distributed according to the unique,
unitarily invariant FS measure on the space
$\mathbb{C}\mathrm{P}^{MN-1}$ of pure states of the composite
system. There is a simple physical motivation for such measures:
they can be used if anything is known about the density matrix,
apart from the dimensionality $M$ of the environment. When $M=1$, we
get the FS measure on the space of pure states. Since the rank of
$\rho$ is limited by $M$, when $M\geq N$ the induced measure covers
the full set of $\mathrm{D}(\mathbb{C}^N)$. Since the pure state
$|\psi\rangle$ is drawn according to the FS measure, the induced
measure is of the product form $P(\rho)=P_{\mathrm{Haar}}({\bf
F}^{(N)})\times P(\Lambda)$. Hence the distribution of the
eigenvectors of $\rho$ is determined by the Haar measure on
$\mathrm{U(N)}$ \cite{IBKZ}.

The measure for the joint probability distribution of spectrum
$\Lambda=\{\lambda_1,\ldots,\lambda_N\}$ of $\rho$ is given by
\cite{KZHJS}
\begin{eqnarray}\label{eq6}
\mathrm{d}\nu_{N,M} (\Lambda)=
C_{N,M}\delta\left(1-\sum^N_{j=1}\lambda_j\right)\prod_{1\leq
i<j\leq
N}(\lambda_i-\lambda_j)^2\prod^N_{j=1}\lambda^{M-N}_j\theta(\lambda_j)\mathrm{d}\lambda_j,
\end{eqnarray}
where $\delta$ is the Dirac delta function, the theta function
$\theta$ ensures that $\rho$ is positive definite, and $C_{N,M}$ is
the normalization constant,
$$
C_{N,M}=
\frac{\Gamma(NM)}{\prod^{N-1}_{j=0}\Gamma(N-j+1)\Gamma(M-j)}.
$$
In particular,  we will consider the case $N=M$ in this paper. In
this scenario, we deal with non-Hermitian square random matrices
characteristic of the {\it Ginibre ensemble} \cite{GINIBRE,MEHTA}
and obtains the Hilbert-Schmidt measure \cite{IBKZ}. Denote
$\mathrm{d\nu_{N,N}=d\nu}$ and $C_N^{\mathrm{HS}}=C_{N,N}$. Thus we
have \cite{LZ5,KZHJS}
\begin{equation}\label{eq7}
\mathrm{d\mu_{HS}(\rho)=d\mu_{Haar}}(U)\times \mathrm{d}\nu(\Lambda)
\end{equation}
for $\rho=U\Lambda U^\dagger$. Here $\mathrm{d\nu(\Lambda)}$ is
given by \cite{LZ5,KZHJS}
\begin{equation}\label{eq8}
\mathrm{d\nu(\Lambda)}=C_N^{\mathrm{HS}}\delta\left(1-\sum_{j=1}^N\lambda_i\right)|\Delta(\lambda)|^2\prod_{j=1}^N
\mathrm{d}\lambda_j,
\end{equation}
where $|\Delta(\lambda)|^2=\prod_{1\leq i<j\leq
N}(\lambda_i-\lambda_j)^2$, and $C^N_{\mathrm{HS}}$ is the normalization constant,
\begin{equation}\label{eq9}
C^N_{\mathrm{HS}}=
\frac{\Gamma(N^2)}{\Gamma(N+1)\prod^{N}_{j=1}\Gamma(j)^2}.
\end{equation}

\vskip0.1in

\noindent {\bf 4. Average skew information-based coherence and its
typicality for random pure states}

\vskip0.1in

We first calculate the average skew information-based coherence for
random pure states.

{\bf Theorem 1} The average skew information-based coherence for a
random pure state $|\psi\rangle\in \mathrm{D}(\mathbb{C}^N)$ is
given by
\begin{equation}\label{eq10}
\mathbb{E}_{\psi}[C_I(\psi)]=\frac{N-1}{N+1}.
\end{equation}

{\bf Proof.} From Eq. (\ref{eq4}), the expected value of the
coherence based on skew information is given by
\begin{equation}\label{eq11}
\mathbb{E}_{\psi}[C_I(\psi)]:=\int
\mathrm{d}\mu(\psi)\left(1-\sum_{k=1}^N |\langle
k|\psi\rangle|^4\right),
\end{equation}
where $\mu$ is a unitarily invariant uniform probability measure.

Take $|\psi\rangle=U|1\rangle$, where $U$ is sampled from the Haar
distribution and $|1\rangle$ is a fixed state. Noting that the Haar
measure is left-invariant, we obtain
\begin{eqnarray*}
\mathbb{E}_{\psi}[C_I(\psi)]
&=&1-\sum_{k=1}^N \int \mathrm{d}\mu(U)|\langle k|U|1\rangle|^4 \nonumber\\
&=&1-N\int \mathrm{d}\mu(U)|U_{11}|^4,
\end{eqnarray*}
where $U_{11}=\langle 1|U|1\rangle$. The distribution of
$|U_{11}|^2$ is given by $(N-1)(1-r)^{N-2}\mathrm{d}r$, where $0\leq
r\leq 1$ \cite{LZ2}. Therefore, we get
\begin{equation}\label{eq12}
\mathbb{E}_{\psi}[C_I(\psi)]=1-N(N-1)\int_{0}^1
r^2(1-r)^{d-2}\mathrm{d}r =1-N(N-1)B(3,N-1),
\end{equation}
where $B(\alpha,\beta)$ is the $\beta$ function
\begin{equation}\label{eq13}
B(\alpha,\beta):=\int_0^1
r^{\alpha-1}(1-r)^{\beta-1}\mathrm{d}r=\frac{\Gamma(\alpha)\Gamma(\beta)}{\Gamma(\alpha+\beta)}.
\end{equation}
Noting that
$$B(3,N-1)=\frac{\Gamma(3)\Gamma(N-1)}{\Gamma(N+2)}=\frac{2}{(N+1)N(N-1)},$$
we obtain from Eq. (\ref{eq12}) the formula (\ref{eq10}). $\Box$

By Theorem 1, it is easy to see that
$\mathbb{E}_{\psi}[C_I(\psi)]=\frac{1}{3}$ for qubit pure states and
$\mathbb{E}_{\psi}[C_I(\psi)]=\frac{1}{2}$ for qutrit pure states.
The limit is $\lim_{N\rightarrow
\infty}\mathbb{E}_{\psi}[C_I(\psi)]=1$ as $N\rightarrow\infty$.

Moreover, it is easy to see that
$(1-\frac{1}{N})-(\frac{N-1}{N+1})<1-\frac{N-1}{N+1}$ for all
integers $N\geq 2$, i.e.,
$\max_{\psi}C_I(\psi)-\mathbb{E}_{\psi}[C_I(\psi)]<\mathbb{E}_{\psi}[C_I(\psi)]-\min_{\psi}C_I(\psi)$,
which means that the average coherence is always closer to the
maximum coherence than the minimum coherence for skew
information-based coherence measure. It can be also found that
$\lim_{N\rightarrow
\infty}(\max_{\psi}C_I(\psi)-\mathbb{E}_{\psi}[C_I(\psi)])=\lim_{N\rightarrow
\infty}\frac{N-1}{N(N+1)}=0$, and $\lim_{N\rightarrow
\infty}\frac{\max_{\psi}C_I(\psi)}{\mathbb{E}_{\psi}[C_I(\psi)]}=\frac{N+1}{N}=1$.
This fact illustrate that for high dimensional quantum systems, the
quantum coherence of a randomly-chosen pure state sampled from the
uniform Haar measure is almost maximal.

Based on the above result, we can further give the following theorem
about the concentration of measure phenomenon for quantum coherence
with respect to random pure states.

{\bf Theorem 2} (Typicality of skew information-based coherence for
random pure states) Let $|\psi\rangle\in \mathrm{D}(\mathbb{C}^N)$
be a random pure state. Then for all $\epsilon>0$, we have
\begin{equation}\label{eq14}
\mathrm{Pr}\left\{\left|C_I(\psi)-\frac{N-1}{N+1}\right|>\epsilon\right\}\leq
2\mathrm{exp}\left(-\frac{N^3\epsilon^2}{72\pi^3\mathrm{ln2}}\right).
\end{equation}

{\bf Proof.} Consider the map $T:|\psi\rangle \rightarrow
T(\psi):=C_I(\psi)$. It follows from Eq. (\ref{eq10}) that
$\mathbb{E}_{\psi}[T(\psi)]=\frac{N-1}{N+1}$. Set $k=2N-1$ in Eq.
(\ref{eq5}). We need to fix the Lipschitz constant $\eta$ for $T$
satisfying $|T(\psi)-T(\phi)|\leq \eta \|\psi-\phi\|_2$. Suppose
that $|\psi\rangle=\sum_{k=1}^N\psi_k|k\rangle$ with
$\sum_{k=1}^N|\psi_k|^2=1$. Denote $p_k=|\psi_k|^2$. Since
$T(\psi)=1-\sum_{k=1}^N |\langle k|\psi\rangle|^4=1-\sum_{k=1}^N
|\psi_k|^4$, we have
\begin{eqnarray}\label{eq15}
\eta^2:=\sup_{\langle\psi|\psi\rangle\leq 1} \nabla T\cdot \nabla T
&=&\sup_{\langle\psi|\psi\rangle\leq 1}\sum_{k=1}^N
(4|\psi_k|^3)^2=\sup_{\langle\psi|\psi\rangle\leq 1}16\sum_{k=1}^N
|\psi_k|^6  \nonumber\\
&=&\sup_{\langle\psi|\psi\rangle\leq 1}16\sum_{k=1}^N p_k^3  \nonumber\\
&=&16N\left(\frac{1}{N}\right)^3=\frac{16}{N^2},
\end{eqnarray}
where the first equality in the last line of Eq. (\ref{eq15}) can be
obtained by using Lagrange multipliers. Therefore, $\eta\leq
\frac{4}{N}$. By definition, we can take $\eta=\frac{4}{N}$ as the
Lipschitz constant. This completes the proof. $\Box$

The inequality (\ref{eq14}) implies that, similar to the relative
entropy of coherence, for large $N$, with high probability, the skew
information-based coherence of $N$-dimensional pure states is
$\frac{N-1}{N+1}$. Namely, most randomly-chosen pure states have
almost $\frac{N-1}{N+1}$ amount of skew information-based coherence.
This is just the so-called concentration of skew information-based
coherence around its expected value, i.e., the typicality of the
skew information-based coherence.

Next, we shall identify a coherent subspace, i.e., a large subspace
of the Hilbert space such that the amount of the skew
information-based coherence for each pure state in this subspace can
be bounded from below almost always by a fixed number that is
arbitrarily close to the typical value of coherence.

{\bf Theorem 3} (Coherent subspaces) Let $\mathcal{H}=\mathbb{C}^N$
be a Hilbert space of dimension $N$. Then for any
$0<\epsilon<\frac{1}{N}$, there exists a subspace
$\mathcal{S}\subset \mathcal{H}$ of dimension
\begin{equation}\label{eq16}
s=\left\lfloor \frac{N^3\epsilon^2-1}{3095(3-\mathrm{ln}\epsilon N)}
\right\rfloor,
\end{equation}
such that all the pure states $|\psi\rangle\in \mathcal{S}$ almost
always satisfy $C_I(\psi)\leq \frac{N-1}{N+1}-\epsilon$. Here
$\lfloor\rfloor$ denotes the floor function.

{\bf Proof.} Let $\mathcal{S}$ be a random $s$-dimensional subspace of $\mathcal{H}$. Let $\mathcal{N}_S$ be an $\epsilon_0$ net for
states on $\mathcal{S}$, where $\epsilon_0=\frac{\epsilon}{4/N}$. It
follows from the definition that $|\mathcal{N}_S|\leq
(5/\epsilon_0)^{2s}$. Identify $\mathcal{S}$ with $U\mathcal{S}_0$,
where $\mathcal{S}_0$ is fixed, and $U$ is a unitary distributed
according to the Haar measure. Endow the net $\mathcal{N}_{S_0}$ on
$\mathcal{S}_0$ and let $\mathcal{N}_S=U\mathcal{N}_{S_0}$. Given
$|\psi\rangle \in \mathcal{S}$, we can choose
$|\tilde{\psi}\rangle\in \mathcal{N}_S$ such that
$\||\psi\rangle-|\tilde{\psi}\rangle\|_2\leq \frac{\epsilon_0}{2}$.
Since $C_I(\psi)$ is a Lipschitz continuous function with the
Lipschitz constant $\eta=\frac{4}{N}$, by the definition of the
$\epsilon_0$ set, we have
$$|C_I(\psi)-C_I(\tilde{\psi})|\leq \eta\||\psi\rangle-|\tilde{\psi}\rangle\|_2
\leq \eta\frac{\epsilon_0}{2}=\epsilon/2.$$

Define $\mathbb{P}=\mathrm{Pr}\{\min_{|\psi\rangle\in
\mathcal{S}}C_I(\psi)<\frac{N-1}{N+1}-\epsilon\}$. From Theorem 2 we have
\begin{eqnarray}\label{eq17}
\mathbb{P}&\leq& \mathrm{Pr}\left\{\min_{|\psi\rangle\in
\mathcal{S}}C_I(\psi)<\frac{N-1}{N+1}-\frac{\epsilon}{2}\right\}\nonumber\\
&\leq&|\mathcal{N}_S|\mathrm{Pr}\left\{C_I(\psi)<\frac{N-1}{N+1}-\frac{\epsilon}{2}\right\}\nonumber\\
&\leq& 2\left(\frac{20}{\epsilon
N}\right)^{2s}\mathrm{exp}\left(-\frac{N^3\epsilon^2}{72\pi^3\mathrm{ln}2}\right).
\end{eqnarray}

If the probability $\mathbb{P}<1$, a subspace with the properties
mentioned in the theorem will exist. This fact holds if
$$s<\frac{N^3\epsilon^2-1}{3095(3-\mathrm{ln}\epsilon N)}.$$
Noting that $\epsilon<\frac{1}{N}$, for $s\geq 2$, we require that
$N\geq 32941$. Therefore, we get $s=\left\lfloor
\frac{N^3\epsilon^2-1}{3095(3-\mathrm{ln}\epsilon N)}
\right\rfloor$. This completes the proof. $\Box$

\vskip0.1in

\noindent {\bf 5. Average skew information-based coherence and its
typicality for random mixed states}

\vskip0.1in

We now turn to the average skew information-based coherence and
its typicality for random mixed quantum states. We first present the following lemma.

{\bf Lemma 1} Denote $|\Delta(\mu)|^2=\prod_{1\leq i<j\leq
N}(\mu_i-\mu_j)^2$. It holds that
\begin{eqnarray}\label{eq18}
&&\int_{\mathbb{R}_+^N}\sqrt{\mu_1\mu_2}\mathrm{exp}\left(-\sum_{j=1}^N
\mu_j\right)|\Delta(\mu)|^2\prod_{j=1}^N \mathrm{d}\mu_j
\nonumber\\
&&=(N-2)!\prod^{N}_{j=1}\Gamma(j)^2\left[\left(\sum_{k=1}^N
I_{kk}^{(\frac{1}{2})}\right)^2-\sum_{k,l=1}^N
\left(I_{kl}^{(\frac{1}{2})}\right)^2\right],
\end{eqnarray}
where $
I_{kl}^{(\frac{1}{2})}=\sum_{r=0}^{\min(k,l)}(-1)^{k+l}\tbinom{\frac{1}{2}}{k-r}\tbinom{\frac{1}{2}}{l-r}\frac{\Gamma(\frac{3}{2}+r)}{r!}.
$

The proof of Lemma 1 is given in Appendix A. Based on the above
lemma, we can give the analytical formula of average skew
information-based coherence for random mixed states in terms of the
dimension $N$.

{\bf Theorem 4} The average skew information-based coherence for a
random mixed state $\rho\in \mathrm{D}(\mathbb{C}^N)$ is given by
\begin{eqnarray}\label{eq19}
\mathbb{E}_{\rho}[C_I(\rho)]&:=&\int_{\mathrm{D}(\mathbb{C}^N)}C_I(\rho)\mathrm{d\mu_{HS}}(\rho)
\nonumber\\
&=&1-\frac{1}{N+1} \left(2+\frac{1}{N^2}\left[\left(\sum_{k=0}^{N-1}
I_{kk}^{(\frac{1}{2})}\right)^2-\sum_{k,l=0}^{N-1}
\left(I_{kl}^{(\frac{1}{2})}\right)^2\right]\right),
\end{eqnarray}
where $\mathrm{d\mu_{HS}}$ is a normalized Hilbert-Schmidt measure,
i.e., $\int_{\mathrm{D}(\mathbb{C}^N})\mathrm{d\mu_{HS}}(\rho)=1$,
and $
I_{kl}^{(\frac{1}{2})}=\sum_{r=0}^{\min(k,l)}(-1)^{k+l}\tbinom{\frac{1}{2}}{k-r}\tbinom{\frac{1}{2}}{l-r}\frac{\Gamma(\frac{3}{2}+r)}{r!}.
$

The proof of Theorem 4 is given in Appendix B. Setting $N=2$ and
$N=3$ in Theorem 4, we obtain the explicit values of the average
coherence for qubit states and qutrit states,
$$\mathbb{E}_{\rho}[C_I(\rho)]=1-\frac{1}{3}\left(2+\frac{3\pi}{16}\right)
=\frac{1}{3}-\frac{\pi}{16}\approx 0.137$$ and
$$\mathbb{E}_{\rho}[C_I(\rho)]=1-\frac{1}{4}\left(2+\frac{103\pi}{256}\right)
=\frac{1}{2}-\frac{103\pi}{1024}\approx 0.184,$$
respectively.

In Figure 1, we plot the average skew information-based coherence
for random mixed states. The $A$-axis shows the value of
$\mathbb{E}_{\rho}[C_I(\rho)]$ given by Eq.(\ref{eq19}). Numerical
calculations show that as the dimension $N$ increases, the
expectation value $\mathbb{E}_{\rho}[C_I(\rho)]$ approaches to a
number which is close to 0.28. Numerical computation shows that
unlike the random pure states, for random mixed states, the average
skew information-based coherence is closer to the minimal coherence
$0$ than the maximum coherence $1-\frac{1}{N}$.

\begin{figure}[ht]\centering
{\begin{minipage}[b]{0.6\linewidth}
\includegraphics[width=0.8\textwidth]{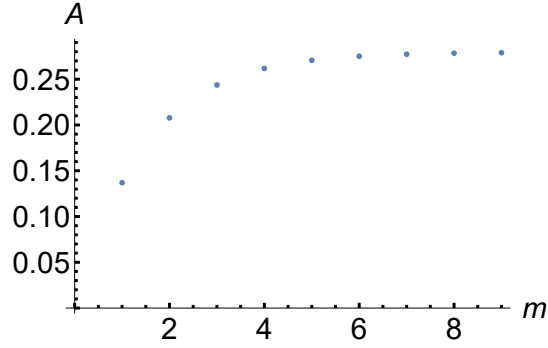}
\end{minipage}}
\caption{The average skew information-based coherence
$A=\mathbb{E}_{\rho}[C_I(\rho)]$ as a function of $N=2^m$.}
\label{fig:Average1}
\end{figure}

Based on the above result, we can similarly discuss the typicality
of quantum coherence $C_I(\rho)$ for random mixed states.

{\bf Theorem 5} (Typicality of skew information-based coherence for
random mixed states) Let $\rho_A\in \mathrm{D}(\mathbb{C}^N)$ be a
random mixed quantum state obtained via partial tracing over a Haar
distributed pure state $|\psi\rangle_{AB}$ in $\mathbb{C}^N\otimes
\mathbb{C}^N$. Then for all $\epsilon>0$, we have
\begin{equation}\label{eq20}
\mathrm{Pr}\left\{\left|C_I(\rho_A)-\mathbb{E}_{\rho}[C_I(\rho_A)]\right|>\epsilon\right\}\leq
2\mathrm{exp}\left(-\frac{N\epsilon^2}{72\pi^3\mathrm{ln2}}\right),
\end{equation}
where $\mathbb{E}_{\rho}[C_I(\rho_A)]$ is given by Eq. (\ref{eq19}).

{\bf Proof.} Define the map $T:\mathbb{S}^{N^2}\mapsto \mathbb{R}$
as $T(\psi_{AB})=C_I(\rho_A)$. Let
$|\psi\rangle_{AB}=\sum_{k,l=1}^N\psi_{kl}|k\rangle_A|l\rangle_B$.
Then $\rho_A=\sum_{k,k'=1}^N p_{kk'}|k\rangle_A\langle k'|$, where
$p_{kk'}=\sum_{l=1}^N \psi_{kl}\overline{\psi_{k'l}}$.
For a bipartitie pure state $|\psi\rangle_{AB}$, it has been shown that
$1-C_I(\psi_{AB})\leq [1-C_I(\rho_A)][1-C_I(\rho_B)]$ \cite{CSY}.
Since $0\leq C_I(\rho_B)\leq 1-\frac{1}{N}$, we have $C_I(\rho_A)\leq
C_I(\psi_{AB})=1-\sum_{k,l=1}^N |\langle k\otimes
l|\psi\rangle|^4=1-\sum_{k,l=1}^N |\psi_{kl}|^4$. Denote
$\tilde{T}(\psi_{AB})=C_I(\psi_{AB})$. Noting that
$p_{kk}=\sum_{l=1}^N |\psi_{kl}|^2$ with $\sum_{k=1}^N p_{kk}=1$, we
have
\begin{eqnarray}\label{eq21}
\eta^2:=\sup_{\langle\psi|\psi\rangle\leq 1} \nabla \tilde{T}\cdot
\nabla \tilde{T} &=&\sup_{\langle\psi|\psi\rangle\leq
1}\sum_{k,l=1}^N (|\psi_{kl}|^3)^2=\sup_{\langle\psi|\psi\rangle\leq
1}16\sum_{k,l=1}^N |\psi_{kl}|^6
\nonumber\\
&=&\sup_{\langle\psi|\psi\rangle\leq 1}16\sum_{k,l=1}^N
(|\psi_{kl}|^2)^3
\nonumber\\
&\leq &\sup_{\langle\psi|\psi\rangle\leq 1}16\left(\sum_{k,l=1}^N
|\psi_{kl}|^2\right)^3=16,
\end{eqnarray}
which implies that $\eta\leq 4$. Now, the Lipschitz constant for $T$
can be obtained in the following way. Suppose that $\sigma_A$ is the
reduced state of another pure state $|\phi\rangle_{AB}$. Without
loss of generality, assume that $C_I(\sigma_A)\leq C_I(\rho_A)$. We
can choose $|\phi\rangle_{AB}$ such that
$C_I(\sigma_A)=C_I(\phi_{AB})$. Then
$$
C_I(\rho_A)-C_I(\sigma_A)\leq C_I(\psi_{AB})-C_I(\phi_{AB})\leq
\eta\||\psi\rangle_{AB}-|\phi\rangle_{AB}\|_2,
$$
Thus the Lipschitz constant of $T$ is bounded by that of $\tilde{T}$
and can be chosen to be $4$. This completes the proof. $\Box$

\vskip0.1in

\noindent {\bf 6. Conclusions and discussions}

\vskip0.1in

We have deduced the explicit formulae for the skew information-based
coherence for both random pure states and random mixed states. It is
found that as $N$ approaches to infinity, the limit of the average
coherence for random pure states is $1$, while this limit for random
mixed states is a positive constant less than $\frac{1}{2}$ by
numerical computation. The average skew information-based coherence
is always closer to the maximum coherence than the minimum coherence
for random pure states, while it is always closer to the minimum
coherence than the maximum coherence for random mixed states, which
demonstrate that for a randomly-chosen state, a quantum pure state
may give rise to more coherence as a resource compared with a
quantum mixed one. This property coincides with the one when
relative entropy of coherence is taken into consideration.

From Eq. (10) it is found that $0\leq
\mathbb{E}_{\psi}[C_I(\psi)]\leq 1$, i.e., the average skew
information-based coherence for a random pure state is always
uniformly bounded, while the average relative entropy of coherence
for a random pure state is $\mathbb{E}_{\psi}[C_r(\psi)]=H_{N}-1$
\cite{LZ2}, which approaches to infinity as the dimension $N$
increases, where $H_N=\sum_{k=1}^N 1/k$ is the $N$th harmonic
number. Unlike a pure state, in \cite{LZ3}, it is shown that the
average relative entropy of coherence for a random mixed state is
$\mathbb{E}_{\rho}[C_r(\rho)]=\frac{N-1}{2N}$. Combining this fact
with the equality given in Eq. (\ref{eq19}), we conclude that in the
mixed state case, the average coherence for skew information-based
coherence and relative entropy of coherence are both uniformly
bounded. Also, it can be seen that
$\mathbb{E}_{\psi}[C_r(\psi)]>\mathbb{E}_{\psi}[C_I(\psi)]$ and
$\mathbb{E}_{\rho}[C_r(\rho)]>\mathbb{E}_{\rho}[C_I(\rho)]$, which
implies that for both a random pure state and a random mixed state,
more coherence as a resource could be generated when the relative
entropy of coherence measure is utilized rather than the skew
information-based one. Moreover, in random pure state case, it is
interesting to note that for skew information-based coherence, the
gap between the maximal coherence and the average coherence is
$1-\frac{1}{N}-\frac{N-1}{N+1}=\frac{N-1}{N(N+1)}>0$, and the limit
approaches to $0$ as $N$ approaches to infinity, while for the
relative entropy of coherence, it is found that this gap
$\mathrm{ln}N-H_{N}+1\gg 0$.

Furthermore, we have shown that the average skew information-based
coherence of pure quantum states (resp. mixed quantum states)
sampled randomly from the uniform Haar measure is typical, i.e., the
probability that the skew information-based coherence of a randomly
chosen pure quantum state (resp. mixed quantum state) is not equal
to the average relative entropy of coherence (within an arbitrarily
small error) is exponentially small in the dimension of the Hilbert
space.

We have also identified a coherent subspace, a large subspace of the
Hilbert space such that the amount of the skew information-based
coherence for each pure state in this subspace can be bounded from
below almost always by a fixed number that is arbitrarily close to
the typical value of coherence. The obtained results in this paper
complement the corresponding results for relative entropy of
coherence, and may shed new light on the study of quantum coherence
from the probabilistic and statistical perspective.

\vskip0.1in

\noindent

\subsubsection*{Acknowledgements}
The authors would like to thank the referees for their valuable
comments, which greatly improved this paper. This work was supported
by National Natural Science Foundation of China (Grant Nos.
11701259, 11971140, 11461045, 11675113), the China Scholarship
Council (Grant No.201806825038), Natural Science Foundation of
Jiangxi Province of China (Grant No. 20202BAB201001), the Key
Project of Beijing Municipal Commission of Education (Grant No.
KZ201810028042), Beijing Natural Science Foundation (Grant No.
Z190005), Natural Science Foundation of Zhejiang Province of China
(Grant No.LY17A010027). This work was completed while Zhaoqi Wu and
Lin Zhang were visiting Max-Planck-Institute for Mathematics in the
Sciences in Germany.

\vskip0.1in

{\bf Appendix A: Proof of Lemma 1}

{\bf Proof of Lemma 1.} Note that $\prod_{1\leq i<j\leq
N}(\mu_i-\mu_j)$ is the classical Vandermonde determinant
$$
\prod_{1\leq i<j\leq N}(\mu_i-\mu_j)= \left|\begin{array}{ccc}
         1&\cdots&1\\
         \mu_1&\cdots&\mu_N\\
         \vdots&\ddots&\vdots\\
         \mu_1^{N-1}&\cdots&\mu_N^{N-1}\\
         \end{array}
\right|.
$$
It can be seen that if $P_0,P_1,\cdots,P_{N-1}$ are polynomials of
respective degrees $0,1,\cdots,N-1$ and respective dominant
coefficients $a_0,a_1,\cdots,a_{N-1}$, one has
$$
\prod_{1\leq i<j\leq N}(\mu_i-\mu_j)= \frac{1}{\prod_{k=0}^{N-1}a_k}
\left|\begin{array}{ccc}
         P_0(\mu_1)&\cdots&P_0(\mu_N)\\
         P_1(\mu_1)&\cdots&P_1(\mu_N)\\
         \vdots&\ddots&\vdots\\
         P_{N-1}(\mu_1)&\cdots&P_{N-1}(\mu_N)\\
         \end{array}
\right|
$$

Now choose $P_k(x)$ to be Laguerre polynomials $L_k(x)$:
$$
L_k(x)=\sum_{j=0}^k (-1)^k\tbinom{k}{k-j}\frac{x^j}{j!}.
$$
Note that $L_k(x)$ have the orthogonality property
\begin{equation}\label{eq22}
\int_0^\infty L_k(x)L_l(x)e^{-x}dx=\delta_{kl},
\end{equation}
and the coefficient of the term with the highest degree is
$a_k=\frac{(-1)^k}{k!}$. We have
\begin{eqnarray}\label{eq23}
\prod_{1\leq i<j\leq N}(\mu_i-\mu_j)^2
&=&\frac{1}{\prod_{k=0}^{N-1}a_k^2} \left|\begin{array}{ccc}
         L_0(\mu_1)&\cdots&L_0(\mu_N)\\
         L_1(\mu_1)&\cdots&L_1(\mu_N)\\
         \vdots&\ddots&\vdots\\
         L_{N-1}(\mu_1)&\cdots&L_{N-1}(\mu_N)\\
         \end{array}
\right|
\nonumber\\
&=&\prod_{k=0}^{N-1}(k!)^2 \sum_{\sigma,\tau\in
S_N}\mathrm{sgn}(\sigma)\mathrm{sgn}(\tau)L_{\sigma(k)-1}(\mu_k)L_{\tau(k)-1}(\mu_k),
\end{eqnarray}
which implies that
\begin{eqnarray*}
&&\int_{\mathbb{R}_+^N}\sqrt{\mu_1\mu_2}\mathrm{exp}\left(-\sum_{j=1}^N
\mu_j\right)|\Delta(\mu)|^2\prod_{j=1}^N \mathrm{d}\mu_j
\\&&=\prod_{k=0}^{N-1}(k!)^2
\sum_{\sigma,\tau\in
S_N}\mathrm{sgn}(\sigma)\mathrm{sgn}(\tau)\left(\int_0^\infty\sqrt{\mu_1}e^{-\mu_1}L_{\sigma(1)-1}(\mu_1)L_{\tau(1)-1}(\mu_1)\mathrm{d}\mu_1\right)
\\&&\left(\int_0^\infty\sqrt{\mu_2}e^{-\mu_2}L_{\sigma(2)-1}(\mu_2)L_{\tau(2)-1}(\mu_1)\mathrm{d}\mu_2\right)\left(\prod_{k=3}^{N}\int_{\mathbb{R}_+^{N-2}}e^{-\mu_k}
L_{\sigma(k)-1}(\mu_k)L_{\tau(k)-1}(\mu_k)\mathrm{d}\mu_k\right),
\end{eqnarray*}
where $S_N$ is the permutation group on $\{1,2,\cdots,N\}$.

Denote $I_{kl}^{(q)}:=\int_0^\infty L_k(x)L_l(x)e^{-x}x^{q}dx$,
where $q>-1$. It holds that \cite{JSR}
\begin{equation}\label{eq24}
I_{kl}^{(q)}=\sum_{r=0}^{\min(k,l)}(-1)^{k+l}\tbinom{q}{k-r}\tbinom{q}{l-r}\frac{\Gamma(q+r+1)}{r!},~~q>-1.
\end{equation}
Note that
$$\int_0^\infty\sqrt{\mu_i}e^{-\mu_i}L_{\sigma(i)-1}(\mu_i)L_{\tau(i)-1}(\mu_i)\mathrm{d}\mu_i=I_{\sigma(i)-1,\tau(i)-1}^{(\frac{1}{2})},~~i=1,2$$
and
$$\int_0^\infty\sqrt{\mu_i}e^{-\mu_i}L_{\sigma(1)-1}(\mu_i)L_{\sigma(2)-1}(\mu_i)\mathrm{d}\mu_i=I_{\sigma(1)-1,\sigma(2)-1}^{(\frac{1}{2})},~~i=1,2.$$

We calculate the integral
$\int_{\mathbb{R}_+^N}\sqrt{\mu_1\mu_2}\mathrm{exp}\left(-\sum_{j=1}^N
\mu_j\right)|\Delta(\mu)|^2\prod_{j=1}^N \mathrm{d}\mu_j$ by
considering the following two cases.

{\bf Case I}: $\sigma=\tau$. Denote
$I=\sum_{k=0}^{N-1}I_{kk}^{(\frac{1}{2})}$, we have
\begin{eqnarray}\label{eq25}
&&\sum_{\sigma,\tau\in
S_N,\sigma=\tau}\mathrm{sgn}(\sigma)\mathrm{sgn}(\tau)\left(\int_0^\infty\sqrt{\mu_1}e^{-\mu_1}L_{\sigma(1)-1}(\mu_1)L_{\tau(1)-1}(\mu_1)\mathrm{d}\mu_1\right)
\nonumber\\
&&\left(\int_0^\infty\sqrt{\mu_2}e^{-\mu_2}L_{\sigma(2)-1}(\mu_2)L_{\tau(2)-1}(\mu_1)\mathrm{d}\mu_2\right)\left(\prod_{k=3}^{N}\int_{\mathbb{R}_+^{N-2}}e^{-\mu_k}
L_{\sigma(k)-1}(\mu_k)L_{\tau(k)-1}(\mu_k)\mathrm{d}\mu_k\right)
\nonumber\\
&&=\sum_{\sigma\in
S_N}I_{\sigma(1)-1,\sigma(1)-1}^{(\frac{1}{2})}I_{\sigma(2)-1,\sigma(2)-1}^{(\frac{1}{2})}
=(N-2)!\sum_{k\neq l}I_{kk}^{(\frac{1}{2})}I_{ll}^{(\frac{1}{2})}
\nonumber\\
&&=(N-2)!\left[\left(\sum_{k=0}^{N-1}I_{kk}^{(\frac{1}{2})}\right)^2-\sum_{k=0}^{N-1}\left(I_{kk}^{(\frac{1}{2})}\right)^2\right].
\end{eqnarray}

{\bf Case II}: $\sigma\neq \tau$. First, note that if there exists
$k_0\in \{3,4,\cdots,N\}$ such that $\sigma(k_0)\neq \tau(k_0)$,
then by Eq. (\ref{eq22}) we have
$$\left(\prod_{k=3}^{N}\int_{\mathbb{R}_+^{N-2}}e^{-\mu_k}
L_{\sigma(k)-1}(\mu_k)L_{\tau(k)-1}(\mu_k)\mathrm{d}\mu_k\right)=0.$$
Thus
$\int_{\mathbb{R}_+^N}\sqrt{\mu_1\mu_2}\mathrm{exp}\left(-\sum_{j=1}^N
\mu_j\right)|\Delta(\mu)|^2\prod_{j=1}^N \mathrm{d}\mu_j=0$.
Otherwise, $\sigma(i)=\tau(i)$ for $i=3,\cdots,N$, which implies
that $\sigma(1)=\tau(2)$ and $\sigma(2)=\tau(1)$, i.e.,
$\tau=\sigma(1 2)$. Then we have
\begin{eqnarray}\label{eq26}
&&\sum_{\sigma,\tau\in S_N,\sigma\neq
\tau}\mathrm{sgn}(\sigma)\mathrm{sgn}(\tau)\left(\int_0^\infty\sqrt{\mu_1}e^{-\mu_1}L_{\sigma(1)-1}(\mu_1)L_{\tau(1)-1}(\mu_1)\mathrm{d}\mu_1\right)
\nonumber\\
&&\left(\int_0^\infty\sqrt{\mu_2}e^{-\mu_2}L_{\sigma(2)-1}(\mu_2)L_{\tau(2)-1}(\mu_1)\mathrm{d}\mu_2\right)\left(\prod_{k=3}^{N}\int_{\mathbb{R}_+^{N-2}}e^{-\mu_k}
L_{\sigma(k)-1}(\mu_k)L_{\tau(k)-1}(\mu_k)\mathrm{d}\mu_k\right)
\nonumber\\
&&=\sum_{\sigma\in
S_N}(-1)I_{\sigma(1)-1,\sigma(2)-1}^{(\frac{1}{2})}I_{\sigma(2)-1,\sigma(1)-1}^{(\frac{1}{2})}
=-(N-2)!\sum_{k\neq l}(I_{kl}^{(\frac{1}{2})})^2.
\end{eqnarray}

Combining Eqs. (\ref{eq25}) and (\ref{eq26}), we have
\begin{eqnarray}\label{eq27}
&&\int_{\mathbb{R}_+^N}\sqrt{\mu_1\mu_2}\mathrm{exp}\left(-\sum_{j=1}^N
\mu_j\right)|\Delta(\mu)|^2\prod_{j=1}^N \mathrm{d}\mu_j
\nonumber\\
&&=\prod_{k=0}^{N-1}(k!)^2\left[(N-2)!\left(\left(\sum_{k=0}^{N-1}
I_{kk}^{(\frac{1}{2})}\right)^2-\sum_{k=0}^{N-1}
\left(I_{kk}^{(\frac{1}{2})}\right)^2\right)-(N-2)!\sum_{k\neq
l}\left(I_{kl}^{(\frac{1}{2})}\right)^2\right]
\nonumber\\
&&=(N-2)!\prod^{N}_{j=1}\Gamma(j)^2\left[\left(\sum_{k=0}^{N-1}
I_{kk}^{(\frac{1}{2})}\right)^2-\sum_{k,l=0}^{N-1}
\left(I_{kl}^{(\frac{1}{2})}\right)^2\right],
\end{eqnarray}
where
$$
I_{kl}^{(\frac{1}{2})}=\sum_{r=0}^{\min(k,l)}(-1)^{k+l}\tbinom{\frac{1}{2}}{k-r}\tbinom{\frac{1}{2}}{l-r}\frac{\Gamma(\frac{3}{2}+r)}{r!}.
$$
$\Box$

{\bf Appendix B: Proof of Theorem 4}

{\bf Proof of Theorem 4.} Since $\mathrm{d\mu_{HS}}$ is a normalized
Hilbert-Schmidt measure, by the definition of $C_I(\rho)$, we have
\begin{eqnarray}\label{eq28}
\int_{\mathrm{D}(\mathbb{C}^N)} C_I(\rho)\mathrm{d\mu_{HS}}(\rho)
&=&\int_{\mathrm{D}(\mathbb{C}^N)} \left[1-\sum_{k=1}^N\langle
k|\sqrt{\rho}|k\rangle ^2\right] \mathrm{d\mu_{HS}}(\rho) \nonumber\\
&=&1-\int_{\mathrm{D}(\mathbb{C}^N)} \sum_{k=1}^N\langle k^{\otimes
2}|\sqrt{\rho}^{\otimes 2}|k^{\otimes 2}\rangle
\mathrm{d\mu_{HS}}(\rho)  \nonumber\\
&=&1-\sum_{k=1}^N \left\langle k^{\otimes
2}\left|\int_{\mathrm{D}(\mathbb{C}^N)}\sqrt{\rho}^{\otimes
2}\mathrm{d\mu_{HS}}(\rho)\right|k^{\otimes 2}\right\rangle.
\end{eqnarray}

It suffices to compute the integral $
\int_{\mathrm{D}(\mathbb{C}^N)}\sqrt{\rho}^{\otimes
2}\mathrm{d\mu_{HS}}(\rho). $ In fact, by the factorization in Eq.
(\ref{eq7}), it follows that
\begin{eqnarray}\label{eq29}
&&\int_{\mathrm{D}(\mathbb{C}^N)}\sqrt{\rho}^{\otimes
2}\mathrm{d\mu_{HS}}(\rho) \nonumber\\
&&=\int \mathrm{d\nu(\Lambda)}\int_{\mathrm{U(N)}}\left[(U\otimes
U)(\sqrt{\Lambda}\otimes \sqrt{\Lambda})(U\otimes
U)^{\dag}\mathrm{d\mu_{Haar}}(U)\right].
\end{eqnarray}
Using the following formula for integral over unitary groups
\cite{LZ7}:
\begin{eqnarray}\label{eq30}
&&\int_{\mathrm{U(N)}}(U\otimes U)A(U\otimes
U)^{\dag}\mathrm{d\mu_{Haar}}(U)
\nonumber\\
&&=\left(\frac{\mathrm{Tr}(A)}{N^2-1}-\frac{\mathrm{Tr}(AF)}{N(N^2-1)}\right)\mathbf{1}_{N^2}
-\left(\frac{\mathrm{Tr}(A)}{N(N^2-1)}-\frac{\mathrm{Tr}(AF)}{N^2-1}\right)F,
\end{eqnarray}
where $A\in M_{N^2}(\mathbb{C})$ and $F$ is the swap operator
defined by $F|ij\rangle=|ji\rangle$ for all $i,j=1,2,\cdots,N$, we
have
\begin{eqnarray}\label{eq31}
\int_{\mathrm{U(N)}}(U\otimes U)(\sqrt{\Lambda}\otimes
\sqrt{\Lambda})(U\otimes U)^{\dag}\mathrm{d\mu_{Haar}}(U)
=\frac{N(\mathrm{Tr}\sqrt{\Lambda})^2-1}{N(N^2-1)}\mathbf{1}_{N^2}+\frac{N-(\mathrm{Tr}\sqrt{\Lambda})^2}{N(N^2-1)}F.
\end{eqnarray}

Noting that
\begin{eqnarray}\label{eq32}
\int(\mathrm{Tr}\sqrt{\Lambda})^2\mathrm{d\nu(\Lambda)} &=&\int
\mathrm{d\nu(\Lambda)}+2\int \sum_{1\leq i<j\leq
N}\sqrt{\lambda_i\lambda_j}\mathrm{d\nu(\Lambda)} \nonumber\\
&=& 1+2\int \sum_{1\leq i<j\leq
N}\sqrt{\lambda_i\lambda_j}\mathrm{d\nu(\Lambda)} \nonumber\\
&=& 1+2C_{\mathrm{HS}}^N\int_{\mathbb{R}_+^N} \sum_{1\leq i<j\leq
N}\sqrt{\lambda_i\lambda_j}\delta\left(1-\sum_{j=1}^N\lambda_j\right)|\Delta(\lambda)|^2\prod_{j=1}^N
\mathrm{d}\lambda_j \nonumber\\
&=& 1+2C_{\mathrm{HS}}^N\tbinom{N}{2}\int_{\mathbb{R}_+^N}
\sqrt{\lambda_1\lambda_2}\delta\left(1-\sum_{j=1}^N\lambda_j\right)|\Delta(\lambda)|^2\prod_{j=1}^N
\mathrm{d}\lambda_j,
\end{eqnarray}
where $C_{\mathrm{HS}}^N$ is given in Eq. (\ref{eq9}), we only need
to calculate
$$
\int_{\mathbb{R}_+^N}\sqrt{\lambda_1\lambda_2}\delta\left(1-\sum_{j=1}^N\lambda_j\right)|\Delta(\lambda)|^2\prod_{j=1}^N
\mathrm{d}\lambda_j.
$$

Denote
$$F(t)=\int_{\mathbb{R}_+^N}\sqrt{\lambda_1\lambda_2}\delta\left(t-\sum_{j=1}^N\lambda_j\right)|\Delta(\lambda)|^2\prod_{j=1}^N
\mathrm{d}\lambda_j.$$ By performing Laplace transform
$(t\rightarrow s)$ of $F(t)$, and letting $\mu_j=s\lambda_j,j=1,2$,
we get
\begin{eqnarray}\label{eq33}
\tilde{F}(s)&=&\int_{\mathbb{R}_+^N}\sqrt{\lambda_1\lambda_2}\mathrm{exp}\left(-s\sum_{j=1}^N
\lambda_j\right)|\Delta(\lambda)|^2\prod_{j=1}^N
\mathrm{d}\lambda_j  \nonumber\\
&=& s^{-(N^2+1)}\int_{\mathbb{R}_+^N}
\sqrt{\mu_1\mu_2}\mathrm{exp}\left(-\sum_{j=1}^N
\mu_j\right)|\Delta(\mu)|^2\prod_{j=1}^N \mathrm{d}\mu_j.
\end{eqnarray}

Utilizing the inverse Laplace transform $(s\rightarrow
t):\mathscr{L}^{-1}(s^{\alpha})=\frac{t^{-\alpha-1}}{\Gamma(-\alpha)}$,
we obtain
\begin{equation}\label{eq34}
F(t)=\frac{t^{N^2}}{\Gamma(N^2+1)}\int_{\mathbb{R}_+^N}\sqrt{\mu_1\mu_2}\mathrm{exp}\left(-\sum_{j=1}^N
\mu_j\right)|\Delta(\mu)|^2\prod_{j=1}^N \mathrm{d}\mu_j.
\end{equation}
Thus
\begin{eqnarray}\label{eq35}
&&\int_{\mathbb{R}_+^N}\sqrt{\lambda_1\lambda_2}\delta\left(1-\sum_{j=1}^N\lambda_j\right)|\Delta(\lambda)|^2\prod_{j=1}^N
\mathrm{d}\lambda_j
\nonumber\\
&&=\frac{1}{\Gamma(N^2+1)}\int_{\mathbb{R}_+^N}\sqrt{\mu_1\mu_2}\mathrm{exp}\left(-\sum_{j=1}^N
\mu_j\right)|\Delta(\mu)|^2\prod_{j=1}^N \mathrm{d}\mu_j.
\end{eqnarray}
Substituting Eq. (\ref{eq18}) into Eq. (\ref{eq35}) yields
\begin{eqnarray}\label{eq36}
&&\int_{\mathbb{R}_+^N}\sqrt{\lambda_1\lambda_2}\delta\left(1-\sum_{j=1}^N\lambda_j\right)|\Delta(\lambda)|^2\prod_{j=1}^N
\mathrm{d}\lambda_j
\nonumber\\
&&=\frac{(N-2)!\prod^{N}_{j=1}\Gamma(j)^2}{\Gamma(N^2+1)}\left[\left(\sum_{k=1}^N
I_{kk}^{(\frac{1}{2})}\right)^2-\sum_{k,l=1}^N
\left(I_{kl}^{(\frac{1}{2})}\right)^2\right],
\end{eqnarray}
which by Eqs. (\ref{eq9}) and (\ref{eq32}) gives rise to
\begin{eqnarray}\label{eq37}
\int(\mathrm{Tr}\sqrt{\Lambda})^2\mathrm{d\nu(\Lambda)}
=1+\frac{1}{N^2}\left[\left(\sum_{k=1}^N
I_{kk}^{(\frac{1}{2})}\right)^2-\sum_{k,l=1}^N
\left(I_{kl}^{(\frac{1}{2})}\right)^2\right].
\end{eqnarray}

Combining Eqs. (\ref{eq29}), (\ref{eq31}) and (\ref{eq37}), we
obtain
\begin{eqnarray*}
&&\int_{\mathrm{D}(\mathbb{C}^N)}\sqrt{\rho}^{\otimes
2}\mathrm{d\mu_{HS}}(\rho)
\\&&=\int \left[\frac{N(\mathrm{Tr}\sqrt{\Lambda})^2-1}{N(N^2-1)}\mathbf{1}_{N^2}+\frac{N-(\mathrm{Tr}\sqrt{\Lambda})^2}{N(N^2-1)}F\right]\mathrm{d\nu(\Lambda)}
\\&&=\frac{N\mathbf{1}_{N^2}-F}{N(N^2-1)}\int(\mathrm{Tr}\sqrt{\Lambda})^2\mathrm{d\nu(\Lambda)}+\frac{NF-\mathbf{1}_{N^2}}{N(N^2-1)}\int \mathrm{d\nu(\Lambda)}
\\&&=\frac{N\mathbf{1}_{N^2}-F}{N(N^2-1)}
\left(1+\frac{1}{N^2}\left[\left(\sum_{k=1}^N
I_{kk}^{(\frac{1}{2})}\right)^2-\sum_{k,l=1}^N
\left(I_{kl}^{(\frac{1}{2})}\right)^2\right]\right)
+\frac{NF-\mathbf{1}_{N^2}}{N(N^2-1)}.
\end{eqnarray*}
Finally, by using the fact that $\sum_{k=1}^N \langle k^{\otimes
2}|F|k^{\otimes 2}\rangle=N,$ we have
$$\sum_{k=1}^N \langle k^{\otimes 2}|N\mathbf{1}_{N^2}-F|k^{\otimes 2}\rangle
=\sum_{k=1}^N \langle k^{\otimes 2}|NF-\mathbf{1}_{N^2}|k^{\otimes
2}\rangle=\frac{N^2-N}{N(N^2-1)}=\frac{1}{N+1}.$$ From Eq.
(\ref{eq28}) we get (\ref{eq19}). $\Box$

\end{document}